\documentstyle[12pt,aasms4,epsf]{article}

\newcommand{\Ms}{M$_{\odot}$}
\newcommand{\kms}{km~s$^{-1}$}

\newcommand{\nv}{N~V $\lambda$1240}
\newcommand{\siiv}{Si~IV $\lambda$1400}
\newcommand{\civ}{C~IV $\lambda$1550}

\newcommand{\heii}{He~II $\lambda$1640}

\newcommand{\siiia}{Si~II $\lambda$1260}
\newcommand{\oi}{O~I/Si~II $\lambda$1303}
\newcommand{\cii}{C~II $\lambda$1335}

\newcommand{\alii}{Al~II $\lambda$1670}

\tighten

\journalid{337}{15 January 1989}
\articleid{11}{14}

\slugcomment{accepted for publication in the ApJL}

\begin{document}

\title{{\it HST} Ultraviolet Spectroscopy of NGC~1741 ---\\
   a Nearby Template for Distant Energetic Starbursts\footnote{Based on
observations with the NASA/ESA {\it Hubble Space Telescope}, obtained at
the Space Telescope Science Institute, which is operated by AURA for NASA
under contract NAS5-26555}}

\author{Peter S. Conti}
\affil{JILA, University of Colorado, Boulder, CO 80309-0440\\
       e-mail: pconti@jila.colorado.edu}

\author{Claus Leitherer}
\affil{Space Telescope Science Institute, 3700 San Martin Drive, Baltimore,
       MD 21218\\
       e-mail: leitherer@stsci.edu}

\and
\author{William D. Vacca}
\affil{Institute of Astronomy, University of Hawaii,
       2680 Woodlawn Drive, Honolulu, HI 96822\\
       e-mail: vacca@athena.ifa.hawaii.edu}

\begin{abstract}
We have obtained a {\it Hubble Space Telescope} ultraviolet image and
spectrum of the nearby Wolf-Rayet galaxy NGC~1741. The spatial
morphology from the Faint Object Camera image is dominated by two main
starburst centers, each being about 100 times as luminous as 30~Doradus.
Both starburst centers are composed of several intense knots of recent star
formation. A Goddard High Resolution Spectrograph spectrum of a portion of
the southern starburst center is consistent with a population of young
stars following a Salpeter IMF for masses above $\sim$15~\Ms\ (lower mass
stars may also be present), and extending up to $\sim$100~\Ms; about $10^4$
O-type stars are inferred from the UV luminosity.

Numerous strong interstellar lines are detected. Although not resolved,
their strength suggests that they are formed in individual bubbles and
shells with velocities up to a few hundred \kms. The red wing of the
Lyman-$\alpha$ absorption profile indicates the presence of several neutral
hydrogen components, one in our own Galaxy and the others at or close to
the distance of NGC~1741. Overall, the stellar and interstellar line
spectrum, as well as the continuum shape of NGC~1741, strongly resembles
star-forming galaxies recently discovered at high redshift.

\end{abstract}


\section{Introduction}

Starburst galaxies are a class of objects experiencing brief but intense
episodes of star formation. Massive OB stars are dominant contributors to
their ultraviolet and optical spectra in the first 10~Myr after the onset
of the burst (e.g., Leitherer et al. 1996). Wolf-Rayet (W-R) galaxies are a
subset of starburst galaxies that show {\em broad} stellar He~II
$\lambda$4686 emission in their spectra (Kunth \& Sargent 1981; Conti
1991). Wolf-Rayet stars are the descendants of the most massive O
stars, which show the products of nuclear burning at their surface due to
mass loss and mixing processes (e.g., Abbott \& Conti 1987). Because only
the most massive stars evolve to form W-R stars, their presence in a
starburst
galaxy might suggest a stellar population with numerous high-mass stars
present (Sargent \& Filippenko 1991), {\it and/or} a rapid formation
timescale, (e.g., Vacca \& Conti 1992, hereafter VC). The census of W-R
stars in the Galaxy and in its OB associations indicates that they have
progenitor masses above
$\sim$40~\Ms\ (Conti et al. 1983; Humphreys, Nichols, \& Massey 1985).
Evolutionary synthesis models of stellar populations containing W-R stars
(e.g., Meynet 1995) predict that about 3 to 6~Myr after the start of the
burst the most massive stars evolve into W-R objects and the ratio of W-R
to O stars increases dramatically. W-R galaxies therefore offer the unique
opportunity to study a burst population whose age is known {\it a priori}.
The determination of the age $t$  and the initial mass function (IMF) of a
starburst can in some cases be degenerate, i.e., the absence of
40~\Ms\ stars could imply either an upper mass limit M$_{\rm up}< 40$~\Ms\
or $t~>$~5~Myr.

NGC~1741 (=~Mrk1089 =~Arp259) was first described as a W-R galaxy by Kunth
\& Schild (1986). It has a highly disturbed optical morphology with two
starburst centers, possibly arising from a galaxy merger, most likely
as a result of interaction with other members of Hickson Compact Group 31
(Hickson, Kindl, \& Auman 1989; Rubin, Hunter \& Ford 1990). NGC~1741 is
one of the most
luminous W-R galaxies --- optically and in the far-UV --- in Conti's (1991)
catalog: $M_B = -20.3$.  VC obtained optical spectrophotometry of NGC~1741
and inferred the presence of about 700 W-R stars in the southern starburst
center (``B'' in
their notation). We selected NGC~1741 as one of the targets for our ongoing
program of {\it Hubble Space Telescope} ({\it HST}) observations of W-R
galaxies. Ultraviolet imaging and spectroscopy have been obtained,
confirming the presence of numerous OB stars in this galaxy. One purpose of
this {\it Letter} is to draw attention to the spectral similarity between
NGC~1741 and starburst galaxies recently discovered at high
redshift, e.g., the ``primeval galaxy candidate'' cB58 (Yee et al. 1996),
the star--forming galaxies at $z > 3$ (Steidel et al. 1996), and the
``lensed'' galaxy at $z = 2.5$ (Ebbels et al. 1996).

\section{Spatial morphology}

We obtained a pre-COSTAR ultraviolet image of NGC~1741 with the Faint
Object Camera (FOC) on board the {\it HST} on 10 March 1993. The
observation was made in the f/96 configuration, in the ``zoomed'' format,
with the F220W filter (effective wavelength of $\sim 2200$~\AA). The total
exposure time was 996~s. The image format consists of $512 \times 1024$
pixels and has a field of view of $\sim 22'' \times 22''$. At NGC~1741's
distance of 51~Mpc (VC; with $H_0$ = 75 km~s$^{-1}$Mpc$^{-1}$), this
corresponds to a plate scale of $5.6$~pc~pixel$^{-1}$, or
250~pc~arcsec$^{-1}$. Reduction and analysis of the image was done in the
same manner as for the W-R galaxy NGC~4214 (Leitherer et al. 1996), and the
reader is referred to that paper for details.

The FOC image is reproduced in Figure~1 (Plate xxx). A preliminary
discussion of this image has been given by Vacca (1994). Two star-forming
centers are clearly resolved into numerous point-like objects, or starburst
knots. The luminosity at 2200~\AA\ of the {\it individual} knots ranges
beyond an order of magnitude larger than that of 30~Doradus. (We will
discuss the photometry along with recently acquired WFPC2 optical images of
NGC~1741 in a subsequent paper). The northeastern center
(J2000 coordinates 5:01:37.76, -4:15:29.0 and extending  over about $2''$)
corresponds to region ``A'' of VC. The southwestern center, overfilling the
Goddard High Resolution Spectrograph (GHRS) aperture, is VC's region ``B''
and is responsible for the W-R features observed in the optical spectrum.

\section{Optical and ultraviolet spectra}

An optical spectrum of NGC~1741B had been obtained and discussed by
VC. Because the spectrum was not shown in that paper, it is included here
for completeness (Figure~2). The long slit spectrum (width $1.5''$, PA
17.5$^{\circ}$) includes all the flux measured by the GHRS aperture, plus
some outside.  The optical spectrum is dominated by strong nebular emission
lines, as expected from the presence of hot, massive stars. The number of
ionizing photons in the Lyman continuum is $2 \times 10^{53}$~s$^{-1}$
(VC). This is more than 2 orders of magnitude higher than observed in
30~Doradus. VC used the $\lambda 4363$ [O~III] line to derive an oxygen
abundance of O/H = $2 \times 10^{-4}$ ($\frac{1}{4} Z_{\odot}$ ), lower
than the 30~Doradus value. The broad emission line at $\lambda$4686 is due
to W-R stars. It is the only directly observed line from hot, massive stars
in the optical range in this galaxy. Weak absorption lines are recognizable
at the wavelength of the higher Balmer series and at the Ca~K line,
suggesting that stars of spectral type late B and cooler contribute to the
optical continuum as well.

The GHRS observations were executed on 17 August 1995. NGC~1741B was
acquired on Side-2 with a 5 by 5 spiral search and centered into the
$1.74''$ Large Science Aperture (LSA). Then a side-switch was performed
for observations at two wavelength settings with grating G140L, one
centered on 1320~\AA\ and the other on 1600~\AA. Wavelength calibrations
were obtained for both grating positions. The reduced spectrum is shown in
Figure~3. A more detailed data analysis will be given subsequently. We will
refer to the portion of NGC~1741B enclosed within the GRHS aperature
(Figure 1) as NGC~1741B1.

The ultraviolet spectra of NGC~1741B1 is similar to knots in other
starburst galaxies observed with {\it HST}, such as NGC~4214 (Leitherer et
al. 1996). However, no starburst galaxy as been observed before in the
ultraviolet at a spectral resolution and signal--to--noise comparable to
our data presented here. The geocoronal Lyman-$\alpha$ was measured at
1216.1~\AA, consistent with the expected wavelength accuracy of this
grating/aperture combination. The narrow interstellar lines of NGC~1741B1
show a small blueshift of about 1~\AA\ in its rest frame (defined from the
optical nebular features). This blueshift is also present in the Galactic
foreground interstellar lines. Most likely, this offset results from a
nonuniform light distribution of NGC~1741B1 in the LSA, which is projected
on 8 science diodes of the detector (Figure~1). Lyman-$\alpha$ has a
FWHM of 4.9~\AA, as expected for a uniform light source filling the LSA
completely. The theoretical FWHM of a spectral line due to a point source
is 0.7~\AA. All interstellar lines have widths $\sim$3~\AA. We
interpret this as a result of the complex structure of NGC~1741B1 partially
filling the LSA, rather than having resolved the interstellar lines.

\section{Stellar population and interstellar medium}

The ultraviolet spectral morphology of starbursts and the spectral
synthesis methods of analyses are described in detail in Leitherer, Robert,
\& Heckman (1995) and Leitherer et al. (1996). The most conspicuous {\em
stellar} lines in NGC~1741B1 are \civ, \siiv, \nv, and \heii. These lines
have broad absorptions and/or emissions due to their origin in stellar
winds from hot stars. The \civ\ and \siiv\ profiles indicate a standard IMF
with a Salpeter slope ($\alpha = 2.35$) in the mass range 15~\Ms\ to
100~\Ms\ (the likely presence of lower mass stars cannot be confirmed from
these $UV$ data). The most massive stars {\em currently} present in the
burst are at least above 60~\Ms, and probably around 100~\Ms. This result
is similar to what is found in other starburst galaxies (Robert, Leitherer,
\& Heckman 1993; Leitherer et al. 1996). The blue edge of \siiv\ requires a
population of evolved O supergiants, which appear between 3~Myr and 6~Myr
after the onset of the burst if there is no ongoing star formation. If star
formation continues, the Si~IV profile is consistent with any age above
3~Myr. The existence of W-R stars suggests an age in the same range ---
independent from the line profile-fitting method. Unfortunately, the
observed \nv\ emission is severely blended with Galactic S~II/Si~II
foreground absorption (see below). The best fit for the age of the
starburst is 4--5 Myr.

The inferred monochromatic luminosity of NGC~1741B1 at 1500~\AA\ is
$L_{1500} = 5.6 \times 10^{39}$~erg~s$^{-1}$~\AA$^{-1}$ ($D = 51$~Mpc;
$E(B-V) = 0.15$~mag, see below). This value implies a number of ionizing
stars roughly consistent with the Lyman continuum photon number of $2
\times 10^{53}$~s$^{-1}$ from NGC~1741B (which is larger in area). The GHRS
aperture of NGC~1741 contains one bright starburst knot (giant H~II
region), some fainter knots and some diffuse background (Figure 1). Part of
the diffuse background may be an artifact of the pre-COSTAR FOC image. Note
that the presence of discrete {\it knots} which dominate
the UV image (Figure 1), strongly suggests that the most recent star
formation process took place in {\it bursts} (see also Conti \& Vacca
1994).

The monochromatic luminosity of NGC~1741B1 at $\lambda$1500~\AA\
corresponds
to $\sim 10^4$ O type stars (following Conti 1996).  We measure a
luminosity in the broad He~II emission line at $\lambda$1640~\AA\ of
$L_{He~II~1640} = 3 \times 10^{39}$erg~s$^{-1}$.
As caibrated by Conti (1996), this corresponds to about 500 WN type
stars.  These star numbers are in reasonable agreement with the values of
$1.7 \times 10^4$ O types and 710 WN types inferred from the optical
spectrum of the larger NGC~1741B optical region by VC.

The {\it optical} continuum of NGC~1741B is several times stronger than
predicted by any model that could reproduce the ultraviolet. We interpret
this as partly due to knots other than B1 in the region (Figure 1) and also
as an indication of an older underlying population having an age of at
least a few ten Myr (our raw {\it WFPC2} optical images indicate such a
background). This population can account for the observed Balmer-line
absorptions and for the low H$\beta$ equivalent width of 80~\AA. Model
predictions for W(H$\beta$) are quite robust (see Leitherer \& Heckman
1995) but fail to agree with the observations by a factor of about 3 in the
case of NGC~1741B (data from VC).

We observe two systems of interstellar lines in NGC~1741B1: one originating
there and the other one within our Galaxy. The strongest interstellar lines
are in NGC~1741B1: \siiia, \oi, \cii, Si~II $\lambda$1526, and \alii, as
well as the interstellar components in \civ, \siiv, and \nv.
Note that \cii\ could also have a stellar contribution from early-B
supergiants. There are weaker Galactic lines (blueshifted by 4000~\kms\ in
Figure~3), most notably \siiia\ (blended with the emission component of
\nv) and \cii. Average equivalent widths of unblended interstellar lines in
NGC~1741B1 are around 2~\AA, which is fairly typical for starburst galaxies
(York et al. 1990).  This value can be used to constrain the kinematics of
the interstellar medium. (For reasons given above, the measured line width
is probably not a good indicator). At 1500~\AA\ this corresponds to a
velocity dispersion $\sigma > 100$~\kms\ if the broadening is due to a
single, virialized gravitational motion, and the implied gravitational
masses would be in excess of $10^{11}$~\Ms, too high for an irregular
galaxy such as NGC~1741. Alternatively the observed equivalent width could
be the result of many unresolved, unsaturated, narrow interstellar lines
over a velocity range of a few hundred \kms. Shells and bubbles around
individual starbursts, or even a large-scale outflow of the ISM in
NGC~1741B1 could be responsible. A similar suggestion for the W-R galaxy
NGC~4214 has been made by Leitherer et al. (1996) in order to explain the
anomalous strength of the interstellar lines in that galaxy in comparison
with individual Galactic stars.

The continuum of NGC~1741B1 is only mildly reddened by dust. Comparison
with
theoretical models using the ultraviolet continuum fitting method of
Leitherer et al. (1996) suggests a total $E(B-V) = 0.15$~mag. The average
Galactic foreground neutral hydrogen column density at the position of
NGC~1741 is $4 \times 10^{20}$~cm$^{-2}$ (Stark et al. 1992), suggesting a
foreground reddening of 0.1~mag, which leaves 0.05 mag for internal
reddening. We dereddened the continuum using $E(B-V) = 0.1$~mag and a
Galactic law, and 0.05~mag with the extragalactic extinction law of Kinney
et al. (1994).

The extinction--corrected continuum shows a pronounced turnover shortward
of 1260~\AA\ --- despite the rising continuum of massive stars. The
turnover is due to the wing of a Lyman-$\alpha$ absorption profile, due
partly to Galactic absorption, and also neutral H in NGC~1741 itself. Test
calculations using line profile--fitting demonstrate that the red
Lyman-$\alpha$ wing can be reproduced by a Galactic neutral hydrogen column
density of the order of $10^{21}$~cm$^{-2}$ and (possibly) three
Lyman-$\alpha$ components at or close to the redshift of NGC~1741B1 with a
total column density of several times $10^{19}$~cm$^{-2}$. The blue
Lyman-$\alpha$ wing has a strong blend due to N~I $\lambda$1200, and a
weaker feature of Si~III$\lambda$1206, and thus cannot be easily modeled.
The Lyman-$\alpha$ column density near NGC~1741B1 is typical of strong
``Lyman limit'' lines along lines of sight to QSOs.

\section{Implications for star-forming galaxies at high redshift}

NGC~1741B1 is made up of a $\sim 10^4$ O type stars concentrated in one
strong star--forming knot and some weaker knots plus background, most of
which collectively originated less than 10 Myr ago. Although the hot stars
dominate the energetics of the UV, and the nebular lines (Figure 2) that
dominate the optical are a direct result of thermal excitation of the
interstellar gas, the Lyman-$\alpha$ line is in {\it absorption}. The
lack of Lyman-$\alpha$ emission is probably due to multiple scattering by
dust within NGC~1741. This result has been noted previously in other
starburst systems (e.g., Hartmann et al. 1988). NGC~1741 is clearly {\it
not} a ``primeval galaxy'', as it shows evidence of an underlying older
population of stars.

Nevertheless, the ultraviolet spectrum of NGC~1741B1 (Figure 3) is
strikingly similar to that of the highly redshifted ($z = 2.72$) ``primeval
galaxy candidate'' cB58 of Yee et al. (1996), to the ``star--forming
galaxies'' at redshifts $z > 3$ of Steidel et al. (1996), and to the
``lensed'' galaxy at $z = 2.5$ (Ebbels et al. 1996). The agreement in
spectral morphology extends both from the perspective of the P Cygni wind
profiles of Si~IV and C~IV, and from the strong interstellar lines of O~I,
C~II, Si~II, and Al~II.  The former lines clearly indicate the presence of
O-type stars, for which we have modeled a burst age of a few Myr in
NGC~1741B1. Yee et al. (1996) derive a slightly older age in their burst
model, but they also suggest that ``continuous'' star formation may be
operating for a somewhat longer time in their galaxy.  The photometry and
spectroscopy of cB58 are currently inconclusive regarding this distinction.
The galaxies of Steidel et al. clearly have young stars present from their
UV spectra, but these authors claim the spectral energy distribution
suggests ``continuous'' star formation operates in their systems.

The interstellar lines in cB58 are about a factor two stronger in
equivalent width, whereas those in the Steidel et al. (1996) galaxies are
similar to NGC~1741B1. We have noted above that the equivalent widths of
the interstellar lines in NGC~1741B1 are {\it likely} dominated by
nonthermal processes and thus cannot be safely used for mass
determinations. This {\it might} be a consideration to be aware of in more
distant galaxies.

A luminosity comparison of NGC~1741B1 with the star--forming galaxies at $z
> 3$ is instructive: Steidel et al. (1996) quote $L_{1500} =
10^{41}h_{50}^{-2}$erg~s$^{-1}$\AA$^{-1}$ for $q_0$ = 0.5, and a little
greater for smaller $q_0$.  The value for NGC~1741B1 is $L_{1500} \sim 1.2
\times 10^{40 }$ {\it in the same units} -- about 10 times fainter.  In our
starburst region, the $L_{1500}$ comes from a bright starburst {\it knot},
and some fainter contributors (see Figure 1).  These represent an
appreciable fraction of the {\it optical} light of the starburst region,
where contributions from older stars are also present. The ``lensed''
galaxy is similar in luminosity to the Steidel et al. objects. The Yee et
al. (1996) galaxy is a factor of 10 (to 100 with an uncertain extinction
correction) brighter than the $z > 3$ galaxies. Although NGC~1741B1 is only
a miniature version of the starburst phenomena exhibited by these more
distant systems, its {\it UV spectrum} closely resembles them.  We offer a
very speculative question: could the {\it burst mode} of star formation we
find in NGC~1741B1 have application to the very earliest stages of galaxy
formation, or are they necessarily completely different? High spatial
resolution imaging of the very distant galaxies would be helpful in
addressing their spatial morphology.  Is the star formation process
``galaxy--wide'' or is it concentrated in ``discrete regions''?  Similarly,
is the process relatively ``continuous'', or is it ``burst--like'', in the
very earliest galaxy formation episodes?

\acknowledgments
Support for this work was provided by NASA grant GO-05900.02-94A
from the Space Telescope Science Institute, which is operated
by the Association of Universities for Research in Astronomy, Inc., under
NASA contract NAS5-26555. WDV has support from the Beatrice Watson Parrent
Foundation. We are very appreciative of stimulating discussions with Erica
Ellingson, and for an advanced copy of the Yee et al. spectra and paper.
We gratefully acknowledge David Bowen's help in computing a series of
damped Lyman-$\alpha$ profiles. We greatly appreciate Mauro Giavalisco's
efforts to draw our attention to star-forming galaxies at high redshift.
Max Pettini pointed out the significance of the equivalent widths of the
interstellar lines in starbursts. This research has made use of the
NASA/IPAC Extragalactic Database (NED) which is operated by the Jet
Propulsion Laboratory, Caltech, under contract with NASA.


\clearpage

\begin{figure}
\plotone{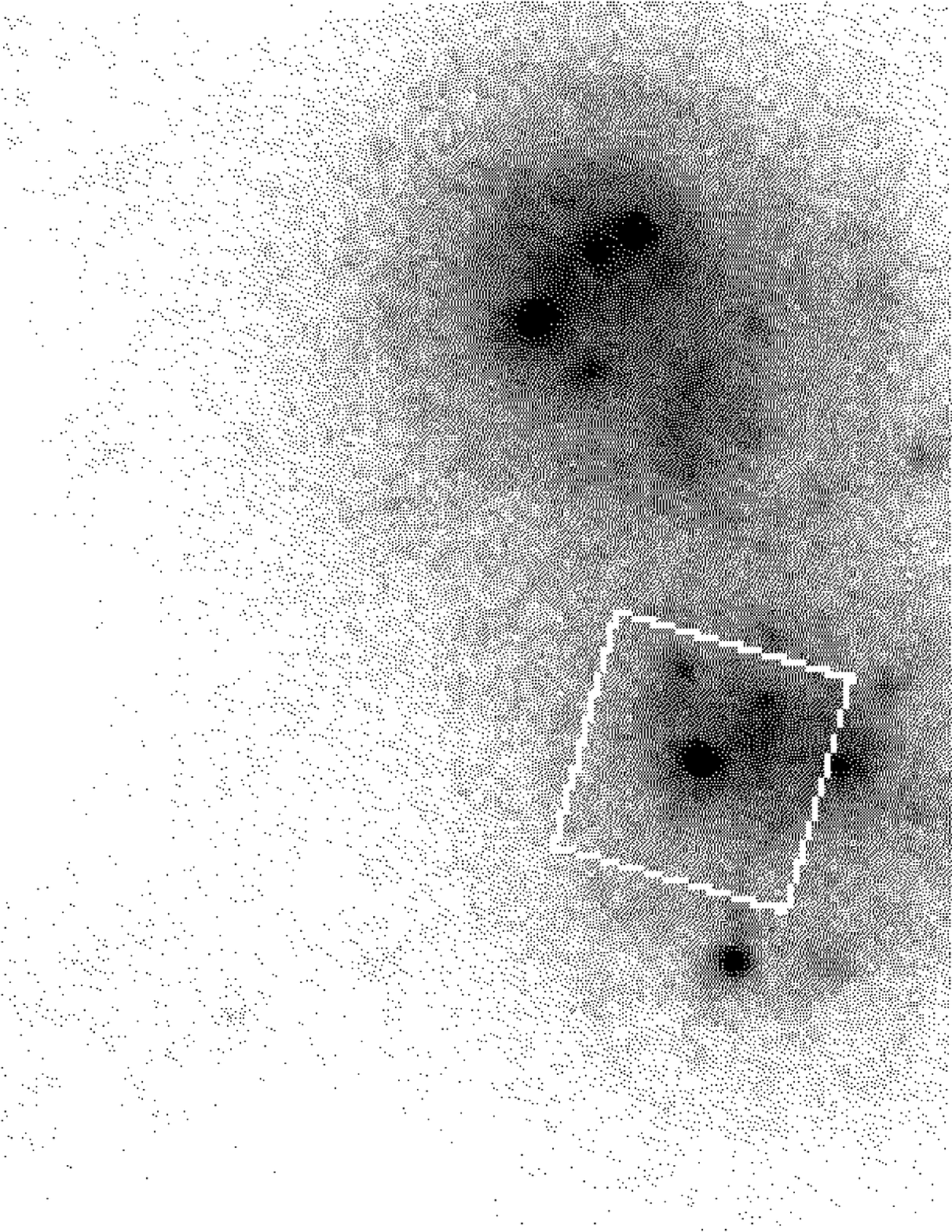}
\caption{Ultraviolet image of NGC~1741 observed with the FOC.
The GHRS aperture (white box) is 1.7$''\times 1.7''$.
North is up, east to the left. NGC~1741A is the upper
starburst region; NGC~1741B, the lower.
}

\end{figure}

\clearpage

\begin{figure}
\plotfiddle{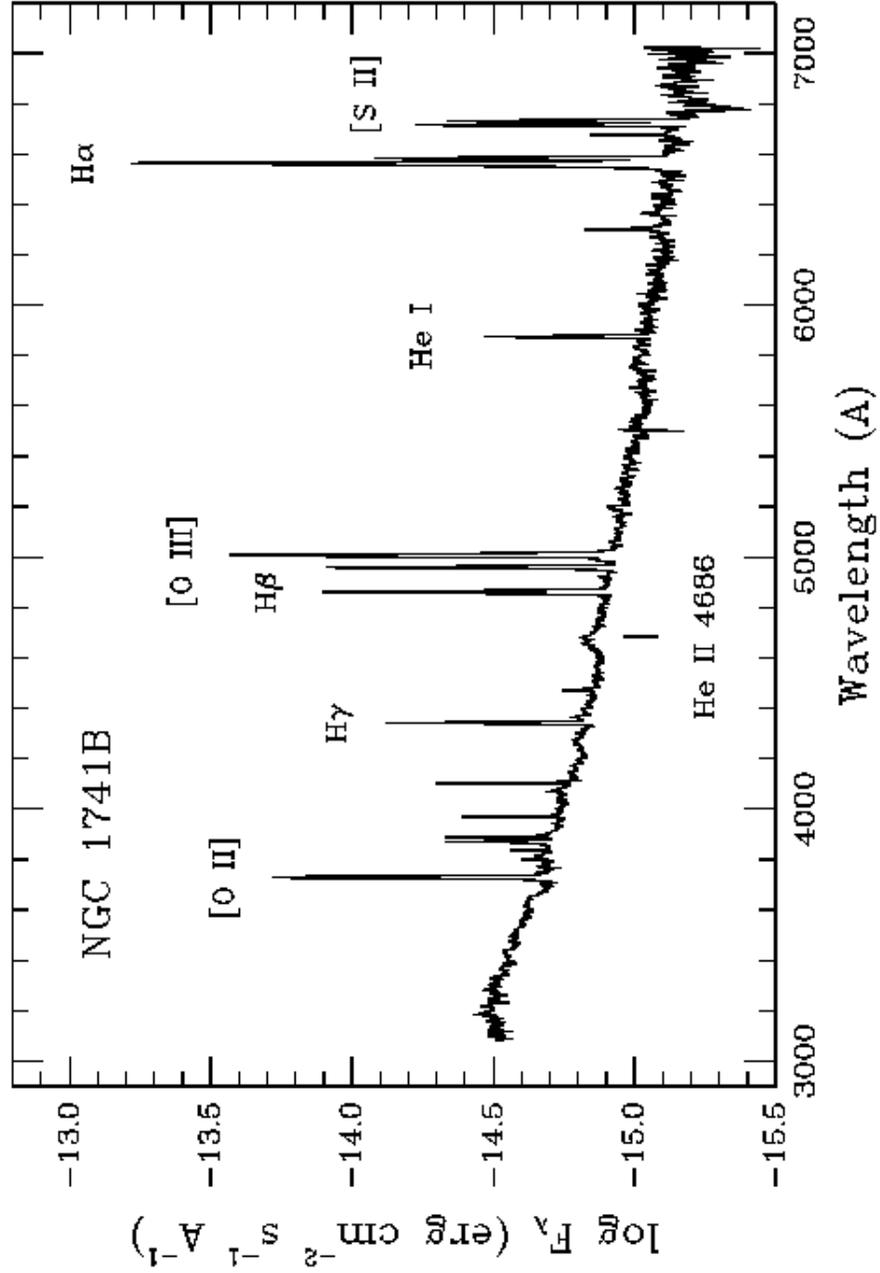}{18cm}{0}{70}{70}{-220}{-10}
\caption{Long slit ($1.5''$ wide) optical spectrum of NGC~1741B, taken at
the 4 m telescope of CTIO and the 2D-Frutti detector (see VC for details).
The wavelength scale is in the rest frame of NGC~1741 (cz~=~4000~\kms).
Note the broad $\lambda$4686 emission due to W-R stars.}
\end{figure}

\clearpage

\begin{figure}
\plotfiddle{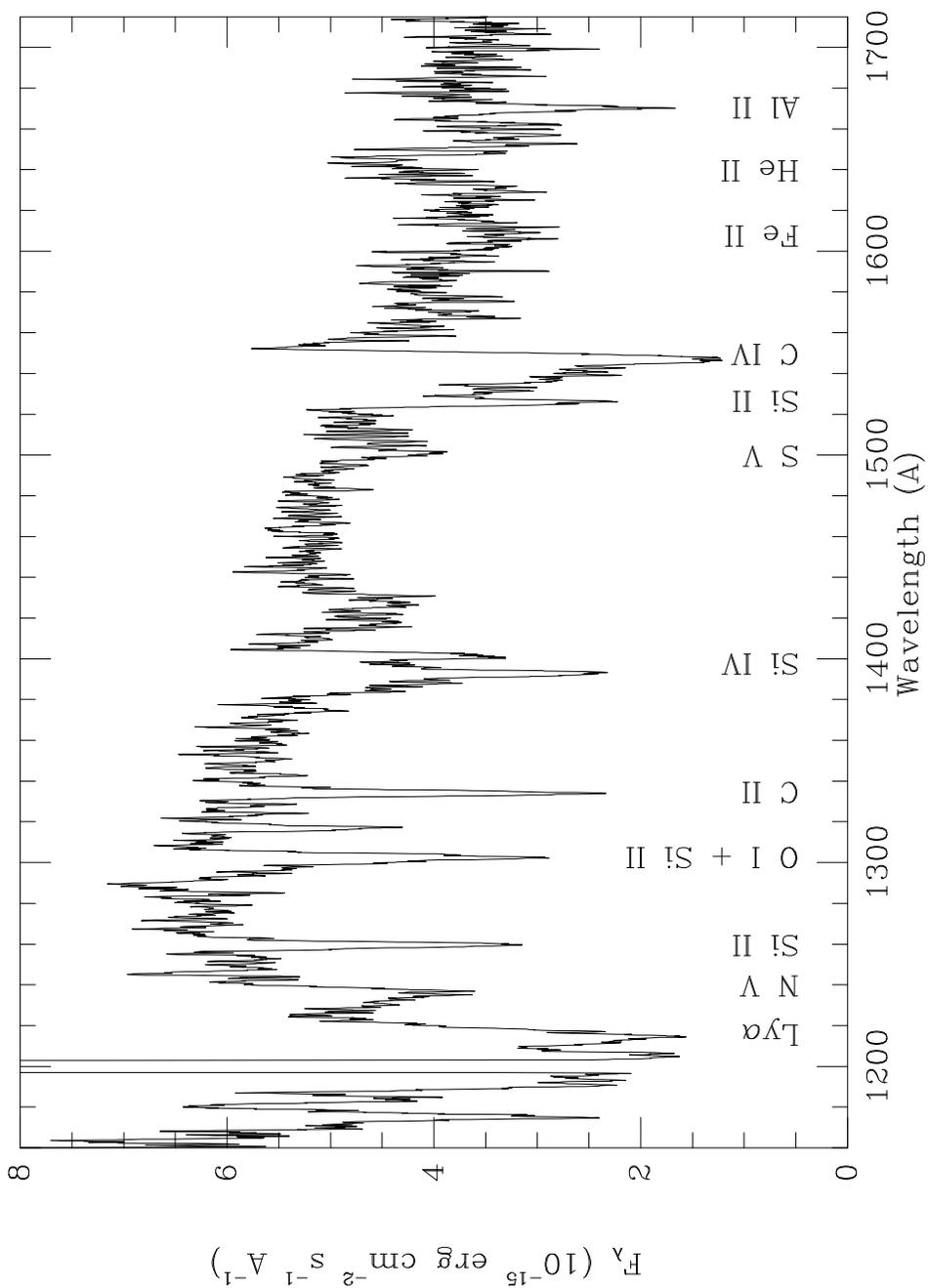}{18cm}{0}{70}{70}{-190}{-10}
\caption{Observed GHRS spectrum of NGC~1741B1 through the
$1.74''$ LSA (see Figure~1). The spectrum was obtained by merging two
spectra centered on wavelength positions of 1320~\AA\ and 1600~\AA.
Exposure times were 6100~s and 12200~s, respectively. A boxcar filter over
5 pixels has been applied, and the spectrum has been blueshifted into the
rest frame of NGC~1741.}
\end{figure}

\clearpage
\end{document}